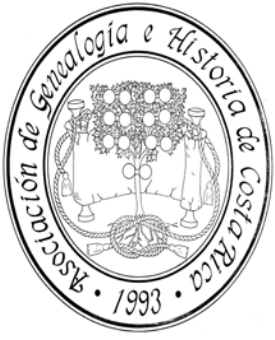

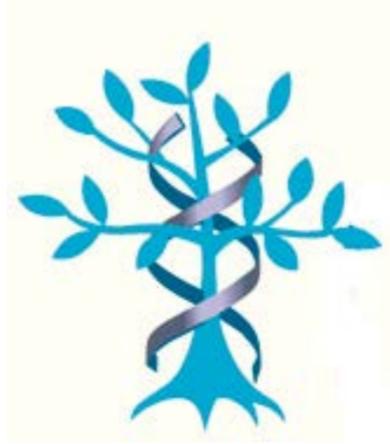

*ASOGEHInforma* es el medio de comunicación interno de la Asociación de Genealogía e Historia de Costa Rica

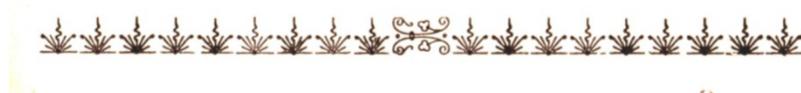

## GENEALOGIAS DE...

Para Emilio Obando Cairol,
por su importante contribución al desarrollo de la genealogía científica en Costa Rica.

### La genealogía mitocondrial de María Mercedes Cairol Antúnez, huella de la inmigración reciente a Costa Rica

Bernal Morera

Miembro de la Asogehi. Dirección para correspondencia: Laboratorio de Genética Evolutiva, Escuela de Ciencias Biológicas, Universidad Nacional, Heredia, Costa Rica; bernal.morera@gmail.com o bernal.morera.brenes@una.cr

El vertiginoso avance de las técnicas de la biología molecular, con apoyo de la antropología, permite en la actualidad detectar procesos demográficos de las poblaciones humanas. Fenómenos migratorios, expansiones poblacionales y mezclas de grupos, entre otros, quedan reflejados en la composición genética de las poblaciones actuales. Por ello el estudio de diferentes marcadores moleculares en muchas poblaciones nos ofrece una visión retrospectiva de la historia de estas, que permite contrastar hipótesis sugeridas por otras disciplinas como la arqueología, la lingüística, la paleontología y la genealogía (Bertanpetit *et al.*, 1999).

Además, ha venido adquiriendo una progresiva popularidad el uso de esta tecnología para profundizar en el conocimiento sobre los ancestros, y de las relaciones de parentesco entre las personas (Meléndez Obando 2004). Iniciativas como el Proyecto Genográfico (National Geographic Society, 2012), la Fundación Sorenson de Genealogía Molecular (SMGF, 2012) y varias compañías privadas que ofrecen servicios para el análisis del ADN mitocondrial (ADNmt) –de herencia materna estricta–, y del cromosoma Y –de herencia exclusivamente paterna–. Así, muchas personas deciden caracterizar sus marcadores genéticos como una iniciativa personal y hacen públicos voluntariamente sus datos genéticos con los objetivos de buscar parientes biológicos y de adquirir conocimiento respecto a su origen remoto.

Nuestro compañero don Emilio Obando Cairol, genealogista costarricense, tenía dichos propósitos cuando contribuyó con sus muestras biológicas a uno de esos proyectos. Además, entre líneas nos manifestó su deseo de compartir pronto tal información con sus parientes. Su abuela, doña María Mercedes Cairol Antúnez, migró de Cuba a Costa Rica, en 1899, donde ha dejado descendencia. De cuyo linaje materno estricto, él es también portador. El objetivo de este estudio de caso es determinar el origen étnico y comparar la información genealógica y genética en torno a este linaje.

**Materiales y métodos**

**Análisis genealógico**. Se construyó un árbol matrilineal estricto (línea uterina) a partir de los datos relevantes para este estudio, publicados por Obando Cairol (2012) a partir de su propia genealogía.

**Caracterización del ADNmt y análisis de los datos**. Las regiones hipervariables (HVRI y HVR2) del genoma mitocondrial fueron analizadas voluntariamente por el propio probado –como se denomina en genética la persona que es objeto de estudio– y los datos del linaje molecular son públicos en la base de datos de ADNmt de la Fundación Sorenson (SMGF, 2012). Él mismo nos proporcionó las pistas para localizarlos y su consentimiento informado para revelarlos. Denominamos a su linaje mitocondrial "Cairol-Antúnez" por los apellidos de sus abuelos maternos, como aparece referido en dicha base de datos. Los resultados genéticos fueron examinamos mediante un abordaje de minería de datos para confirmar el haplogrupo al cual pertenecen.

Para esclarecer el origen geográfico del linaje observado, comparamos los datos con secuencias publicadas según se describió previamente (Morera y Meléndez Obando 2009, Morera *et al.* 2012). Confeccionamos además una red filogenética mediante el algoritmo de Uniones Medias (Median Joint) con el programa NETWORK 4.6.1.0 (Fluxus Technology, 2012). Con tal propósito comparamos este linaje mitocondrial observado en la región control HVR1, con secuencias europeas del linaje H; indígenas americanas de linajes A, B, C y D; y africanas L2, las cuales también han sido encontradas en costarricenses.

**Resultados**

En el Cuadro genealógico N°1 se presenta la genealogía matrilineal de 4 generaciones documentadas desde doña Ángela Antúnez hasta uno de sus descendientes costarricenses. De acuerdo con los registros documentales, ella era de origen étnico catalán español (Obando Cairol 2009).

El linaje de ADNmt observado en los descendientes costarricenses de doña Ángela Antúnez corresponde a la Secuencia de Referencia de Cambridge (CRS) descrita por Anderson *et al.* (1981), y pertenece al haplotipo (o linaje) H que tiene un origen filogenético europeo. Esto se ilustra en la Figura 1, donde la red de relaciones filogenéticas muestra la estrecha correspondencia del linaje mitocondrial "Cairol- Antúnez" con los linajes europeos del grupo H. Por lo tanto, los resultados genéticos y documentales respecto al origen étnico de esta matriarca costarricense concuerdan entre sí, según lo esperado. En la región I este ADNmt reciente presenta solo una base [HVRI = 16519 C] de diferencia respecto a la secuencia de referencia (CRS, círculo marcado en la figura 1). En tanto que en la región II presenta tres bases de diferencia [HVRII = 263 G, 309.1 C y 315.1 C]. Por otra parte, es claramente diferente de las secuencias de los linajes indígenas americanos A, B, C y D, y del linaje africano L2. Todos estos han sido detectados en costarricense contemporáneos (Figura 1).

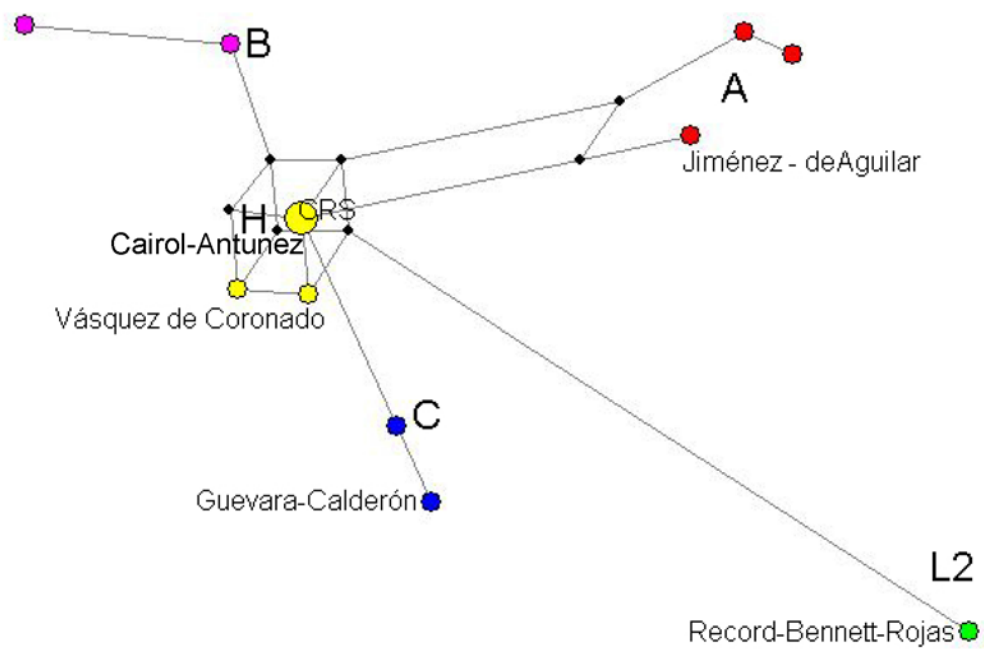

Figura. 1. Red de relaciones genéticas de uniones medias que compara las semejanzas del linaje mitocondrial "Cairol- Antúnez" con otros portadores de secuencias europeas de haplogrupo H (en amarillo), que incluye la Secuencia de Referencia de Cambridge (CRS) indicada en el nodo central, y el linaje "Vásquez de Coronado" (Morera y Meléndez Obando 2012). Para comparación se incluye las secuencias de origen indígena americana de los haplogrupos: A (en rojo) que incluye al linaje "Jiménez" (Morera *et al.* 2012) –"de Aguilar" (Morera y Meléndez Obando 2010), B (en violeta), C (en azul) que incluye el linaje "Guevara-Calderón" (Morera y Meléndez Obando 2009), y D (en azul); y la secuencia de origen africano del haplogrupo L2 (en verde), encontrada en el linaje "Record-Bennett-Rojas" (Ovares Ramírez y Morera 2009) . Todos estos observados previamente en individuos costarricenses. Secuencias no encontradas aún aparecen como un punto en negro.

Otra observación interesante, es que este linaje presenta tan solo una diferencia en la región HVRII [309.1 C], con respecto al linaje previamente estudiado, y que fue observado en los costarricenses descendientes en forma directa de doña Andrea Vásquez de Coronado (Morera y Meléndez Obando, 2012).

**Discusión**

Dado que los haplotipos de ADNmt pueden ser asignados a grupos de secuencias específicas de un continente, pudimos establecer el origen geográfico del linaje de ADNmt presente en la genealogía estudiada. Este coincide cercanamente con la Secuencia de Referencia de Cambridge (CRS). El haplogrupo H tiene alrededor del 40 por ciento del patrimonio genético de algunas ciudades mediterráneas (National Geographic Society, 2012). En general, presenta sus frecuencias más altas en la población de la Península Ibérica (Simoni *et al.*, 2000), tanto como el 19,3% entre españoles (Bertranpetit *et al.* 1995, Côrte-Real *et al.*, 1996, Pinto *et al.*, 1996, Salas *et al.*, 1998) y el 25,9% entre portugueses (Côrte-Real *et al.*, 1996).

Así, podría parecernos obvio o incluso poco llamativo el encontrar otro haplogrupo H, justo en un linaje documental europeo reciente. Por el contrario, este hecho si tiene gran significancia desde una perspectiva biológica, como veremos.

En primer lugar, y en forma semejante a como se ha observado en Quebec, la llegada de mujeres inmigrantes tardías aportó distintos linajes de ADNmit, que junto con la mezcla contribuyeron a enriquecer la diversidad y la regionalización del acervo genético de aquella población francófona de Canadá (Moreau *et al.*, 2009). Un impacto equivalente se esperaría haya ocurrido en la población de Costa Rica.

En segundo lugar, nos comenta Obando Cairol (2009): "lo bello para mí es que la migración de mi abuela Mercedes, distinta a las migraciones usuales que se originan mayormente en situaciones políticas, económicas o desastres naturales, emigró a Costa Rica por amor y gracias a ello nacieron sus descendientes. Era, sin duda, una mujer de armas tomar, industriosa y de carácter, pues ella solita crió a sus hijas en una tierra extraña". Y desde una perspectiva humana, esta sin duda es una historia hermosa.

No obstante, pareciera que es asimismo un ejemplo de un mecanismo migratorio de enorme trascendencia. De hecho la migración femenina ha desempeñado un papel significativo debido a los matrimonios patrilocales practicados en la mayoría de las sociedades, donde las mujeres se trasladan al lugar de origen del esposo (Bertanpetit *et al.*, 1999). Y en este caso si los papeles matrimoniales fueron o no firmados resulta irrelevante, pues la historia de amor culminó como se esperaría, con creciente descendencia.

Estas prácticas sociales quedan reflejadas en los linajes femeninos o masculinos y los avances tecnológicos de la genética molecular nos permiten ponerlas de manifiesto (Perego *et al.* 2005). Un buen ejemplo de esto fue observado por Comas *et al.*, (1998) al estudiar las poblaciones del Asia Central, a lo largo de la legendaria ruta de la seda. Ellos encontraron que el ADNmit presenta una gran homogeneidad entre las diversas poblaciones de aquella región, no detectándose diferencias entre ellas. Contrariamente, el análisis de marcadores genéticos del cromosoma Y muestra resultados diferentes, con un grado muy elevado de heterogeneidad por el lado masculino. Estas discrepancias entre ambos hallazgos sugiere un patrón de migración diferencial entre hombres y mujeres, con una mayor tasa de migración de las mujeres.

Por mucho tiempo consideramos que los procesos migratorios en la historia de las poblaciones humanas se basaban en movimientos predominantemente masculinos debidos a conquistas bélicas, exploraciones, etc. Sin embargo, la migración femenina en pequeña escala ha resultado ser un asunto similar o mayor importancia en la evolución humana (Bertanpetit *et al.*, 1999).

Otro aspecto que amerita ser discutido es la enorme semejanza encontrada entre los linajes "Cairol- Antúnez" y "Vásquez de Coronado" (Morera y Meléndez Obando 2012), los cuales presentan solo una base de diferencia [HVRII = 309.1 C]. Dicha similitud no es casual, sino implica que ambos linajes poseen una mujer ancestro en común. La explicación alternativa, que el primer linaje desciende del segundo, queda descartada tajantemente gracias al conocimiento de ambas genealogías. Ya que doña Ángela Antúnez (muerta hacia 1881) no es descendiente de doña Andrea Vásquez de Coronado (h. 1580 - 1657).

Sin embargo, dicha antecesora en común más reciente de ambas matronas no hay que buscarla en la época de la colonia española en Centroamérica, sino en la lejana última era del hielo, hace unos 10.000 a 15.000 años (Pereira *et al.*, 2005, Álvarez-Iglesias *et al.*, 2009). En aquellos tiempos difíciles, los primitivos europeos se vieron obligados a refugiarse en los sitios con climas más benevolentes. Y fue así como el haplogrupo H sobrevivió en la Península Ibérica.

Finalmente, durante los últimos 500 años, los ancestros maternos de portadores del haplogrupo H, se expandieron hacia el oeste, atravesando el océano Atlántico en barcos de velas durante la conquista y colonización europea del Nuevo Mundo. Tal fue el caso del linaje "Vásquez de Coronado". Más tarde, otros portadores de las líneas mitocondriales de los aquí estudiados, como aquellos del linaje "Cairol- Antúnez", los seguirían en barcos de vapor, que también los traerían al continente americano en su busca de una vida mejor. Y como sabiamente concluiría don Emilio, hasta Costa Rica en busca del amor.

**Cuadro Genealógico N°1**

Genealogía matrilineal de Emilio Gerardo Obando Cairol.

(1)  **Emilio Gerardo Obando Cairol.**
n. San José 20 marzo 1946. b. Iglesia Nuestra Señora de las Mercedes, 1 junio 1946. c. 29 octubre 1967, c. Abigail Mathieu Araya (h. de Wilfridio Mathieu Agüero y Trinidad Araya Azofeifa), en Iglesia Medalla Milagrosa en Barrio Cuba, SJ. Con descendencia.

(3) **Hortensia Matías de la Cruz Cairol Cairol** (recte Odio Cairol).
n. 14 setiembre 1903 en Los Ángeles de Cartago, b. San Rafaél, Oreamuno, Cartago 7-octubre-1903. m. 14 mayo 1979. [Su linaje materno continua a través de los descendientes de su hija Julia María Cairol Cairol (n. SJ 28-12-1929) cc. Edwin Herrera Altamirano. Este llega ahora hasta la sexta generación en Costa Rica].

cc.
Ramón Emigdio de Jesús (Emilio) Obando Bonilla.
n. SJ 10-junio-1870, m. 27 setiembre 1951. (h. José Tomás Obando Alpízar y Feliciana de la Rosa Bonilla Elizondo).

(7). **María Mercedes Cairol Antúnez**.
n. 24 dic. 1879, Barcelona, Cataluña, España. Emigró a Santiago de Cuba en la década de 1880 con su padre y hermano. Emigra a Costa Rica 9 julio 1899. m. 29 octubre 1940 SJ.
Con descendencia en Costa Rica

c.
Prudencio Odio Giró
n. abril 1868, Santiago de Cuba. m. 17 agosto 1940 en Costa Rica
(h. José María Odio Boix y Lucia Giró Giró)

(15). **Ángela Antúnez**
n. Cataluña. m.h. 1881 Barcelona, Cataluña, España.

cc.
Ramón Cairol
n. Cataluña. Emigró a Santiago de Cuba en la década de 1880 con sus hijos
y su segunda esposa, Josefa Prat.

Fuente: Obando Cairol (2009, 2012).